\documentclass[onecolumn,preprint,dvips]{aastex}
\usepackage{ textcomp }
\usepackage[]{natbib}

\newcommand{\numtot}{709}
\newcommand{\numeight}{130}
\newcommand{\numNOH}{19}
\newcommand{\numPM}{286}
\newcommand{\numC}{115}
\newcommand{\numI}{499}
\newcommand{\numFH}{103}
\newcommand{\numFB}{58}
\newcommand{\numH}{49}
\newcommand{\numtotirac}{626}
\newcommand{\numnewhere}{660}
\newcommand{\numnoid}{581}
\newcommand{\numInnerG}{663}

\newcommand{\etal}{et al.}
\newcommand{\sst}{$Spitzer~Space~Telescope$}
\newcommand{\av}{$A_{\rm V}$}
\newcommand{\hii}{\ion{H}{2}}
\newcommand{\HII}{\ion{H}{2}}
\newcommand{\kms}{km~s$^{-1}$}
\newcommand{\msun}{M$_\odot$}

\begin{document}

\author{Henry~A.~Kobulnicky\altaffilmark{1}} 
\author{William T. Chick\altaffilmark{1}}
\author{Danielle P. Schurhammer\altaffilmark{1}}
\author{Julian E. Andrews\altaffilmark{1,2}}
\author{Matthew S. Povich\altaffilmark{2}}
\author{Stephan A. Munari\altaffilmark{1}}
\author{Grace M. Olivier\altaffilmark{1,3}}
\author{Rebecca L. Sorber\altaffilmark{1,4}}
\author{Heather N. Wernke\altaffilmark{1,5}}
\author{Daniel A. Dale\altaffilmark{1}}
\author{Don M. Dixon\altaffilmark{2}}

\altaffiltext{1}{Dept. of Physics \& Astronomy, University 
of Wyoming, Laramie, WY 82070, USA }

\altaffiltext{2}{Department of Physics \& Astronomy, California State Polytechnic
 University, 3801 West Temple Avenue, 
 Pomona, CA 91768, USA}
\altaffiltext{3}{Department of Astronomy, Case Western Reserve University, 
 Cleveland, OH 44106, USA}
\altaffiltext{4}{Front Range Community College, 3645 W. 112th Avenue, 
 Westminster, CO 80031, USA}
\altaffiltext{5}{Department of Physical Sciences, 
 Embry-Riddle Aeronautical University, Daytona Beach, FL 32114, USA}

\slugcomment{Accepted for publication in the Astrophysical Journal}

\title{A COMPREHENSIVE SEARCH FOR STELLAR BOWSHOCK NEBULAE IN THE MILKY WAY: 
A CATALOG OF \numtot\ MID-INFRARED SELECTED CANDIDATES }
\begin{abstract}  

We identify \numtot\ arc-shaped mid-infrared nebula in 24
$\mu$m {\it Spitzer Space Telescope} or 22 $\mu$m {\it Wide
Field Infrared Explorer}  surveys of the Galactic Plane as
probable dusty interstellar bowshocks powered by early-type
stars.   About 20\%  are visible at 8 $\mu$m or shorter
mid-infrared wavelengths as well. The vast majority
(\numnewhere) have no previous identification in the
literature.   These extended infrared sources are  strongly
concentrated near Galactic mid-Plane with an angular scale
height of $\sim$0.6\degr.  All host a symmetrically placed
star implicated as the source of a stellar wind sweeping up
interstellar material. These are candidate ``runaway'' stars
potentially having  high velocities in the reference frame
of the local medium.  Among the \numPM\ objects with
measured proper motions, we find an unambiguous excess 
having velocity vectors aligned with the infrared morphology
--- kinematic evidence  that many of these are ``runaway''
stars with large peculiar motions responsible for the
bowshock signature.  We discuss a population  of ``in-situ''
bowshocks ($\sim$\numFH\ objects) that face giant \hii\
regions where the relative motions between the star and ISM
may be caused by bulk  outflows from an overpressured
bubble.   We also identify $\sim$\numFB\ objects that face 8
$\mu$m bright-rimmed clouds and  apparently constitute a
sub-class of in-situ  bowshocks where the stellar wind
interacts with a photo-evaporative flow from an eroding
molecular cloud interface (i.e., ``PEF bowshocks'').   
Orientations of the arcuate nebulae exhibit a correlation
over small angular scales, indicating that external
influences such as \HII\ regions are responsible for
producing some bowshock nebulae.    However, the vast
majority of this sample appear to be isolated (\numI\
objects) from obvious external influences.  

\end{abstract}


\section{Introduction}

Stellar bowshock nebulae are arcuate structures created by
the interaction between stellar winds and the surrounding
interstellar medium (ISM) where the relative velocity
between the two is supersonic \citep{vanBuren90, maclow91}.
A shock front forms at the interface of the high speed 
stellar wind and the ambient
interstellar medium. The swept up interstellar dust and gas
ahead of the high-velocity star forms an arc-like feature
preceding the peculiar motion of the star. This material is
most clearly visible in the mid-infrared and sometimes in
optical emission lines such as H$\alpha$ and \ion{O}{3}
\citep{brown05,Meyer14, Meyer16}. Most bowshock nebulae are
observed around high-velocity massive stars, but they have
also been identified preceding high-velocity pulsars
\citep{wang13}, red supergiants \citep{Noriega97}, and
associated with proplyds in the Orion Nebula
\citep{Bally98}. \citet{Wilkin96} derived an exact analytic
solution for this interaction by balancing the ISM pressure
and ram pressure of the stellar winds. Based on this model,
\citet{comeron98,Meyer14,Meyer16} used computer simulations
to demonstrate that stable bowshocks can be formed for a
variety of interaction scenarios over a range of relative
velocities, ISM densities, and stellar wind momentum fluxes.

A significant proportion (10$\pm$25\%) of early-type  stars
reside outside of stellar clusters \citet{gies86}. Because
of their short main-sequence lifetimes these stars, known as
runaway stars, must be moving at high peculiar velocities,
typically  \textgreater30 \kms\ \citep{gies86}.
\citet{stone91} found that O and B type stars (OB stars)
comprise 50\% of all runaway stars, and these OB runaways
are dominated by O type stars (12:1).  Two general scenarios
that generate runaway stars have been proposed: close
dynamical interactions of either single stars
\citep{poveda67} or binary-binary interactions
\citep{spitzer80}, and ejection from close binary systems
when the more evolved member undergoes a core-collapse
supernova \citep[CCSN;][]{zwicky57, blaauw61}.

\citet{poveda67} analytically simulated the dynamical
interactions of small clusters of 5--6 50 \msun\ stars and
found that 2--17\% of these stars were ejected from the
cluster with velocities in excess of 35 \kms.
\citet{kroupa01} concluded that dynamical few-body ejections
of OB stars likely occur in the first $\lesssim$1 Myr of the
cluster's history before the radiation of the OB stars
expels the mass-dominant gas component of the cluster,
causing the cluster to expand.  \citet{leonard91} simulated
binary-binary gravitational interactions and found a maximum
ejection velocity of 700 \kms\ for 60 \msun\ stars and 1400
\kms\ for 1--4 \msun\ stars.  Though the initial
multiplicity fraction of OB stars is unknown, studies have
shown that the fraction is likely higher than 60\% in
stellar clusters, meaning that at least 75--90\% of massive
stars are in a multiple system and have the potential to
participate in few-body interactions 
\citep{garcia01,sana12,kiminki12,kobul14}. 

\citet{zwicky57} first proposed that stars that form in
close binary systems could be ejected when one of the
components explodes as a supernova. \citet{zheng15}
hydrodynamically simulated the effects of asymmetric CCSN in
close binary systems and determined that the less-evolved
companion could survive with minimal disruption and be
ejected at runaway velocities. \citet{fryer98} used Monte
Carlo simulations to determine the necessary impulse to
account for the observed velocity distribution of pulsars
and found by extension that the necessary forces could
accelerate the surviving OB companions to velocities up to
100 km s$^{-1}$. \citet{tauris15} expanded upon this work
with new simulations and found that early-type B stars (10
\msun) could be accelerated up to $\sim$320 \kms\ under
ideal conditions with increasing maximum velocities for
lower mass stars (up to $\sim$1050 \kms\ for a 0.90 \msun\
star).

It is likely that both few-body interactions and supernovae
in binary systems combine to generate the observed
population of runaway stars. Using Hipparcos proper motions
\citet{hoogerwerf00} traced the motions of two known runaway
stars, AE Aur and $\mu$ Col (O9.5V and O9.5B/B0, moving in
opposite directions at 100 \kms) to a common origin
$\sim$2.5 million years ago near the location of the binary
pair $\iota$ Ori. They proposed that the two runaways may
have been ejected in a binary-binary interaction.
\citet{Gvaramadze13} identified two O-type runaways that may
have been ejected from the star cluster NGC~3603 in a single
star-binary interaction when the single star captured one
binary member and ejected the other with both systems being
accelerated to high peculiar velocities relative to the
cluster. The new binary system later merged into the single
'blue straggler' star observed. \citet{hoogerwerf00} used
Hipparcos proper motions to trace $\zeta$ Oph and the pulsar
J1932+1059 back to  the same region of the Upper Scorpius
star forming region $\sim$1 Myr ago, which could indicate
that the two were once members of a common binary prior to
the supernova creating the pulsar. \citet{hoogerwerf01}
identified eight additional binary-binary and 16
binary-supernova ejection candidates by extrapolating their
proper motions. A fraction of the ejected stars in the
\citet{zheng15} simulations accreted material from the CCSN,
and these simulations predict that atmospheric chemical
enrichments may be observable in the runaway stars if mixing
is inefficient.  Several runaway stars
\citep{blaauw93,Gvaramadze09} have been observed with
significant $\alpha$-element enrichment, consistent with
this prediction. 

Hybrids of these ejection scenarios may also be
possible. \citet{pflamm10} suggested that a two-step
scenario involving a dynamically ejected binary pair could
 accelerate a single runaway star following a CCSN.
Because of the two kicks accelerating the star in this
scenario, the star's observed proper motion vector may not
point back to a cluster. \citet{wit05} examined a sample of
known runaway stars and found that 4$\pm$2\% do not have
proper motions that can be extrapolated back to a known
cluster or association.

Hypervelocity stars (HVSs) are
a special class of runaway stars with extremely high
space velocities, typically defined as being greater
than 400 \kms\ \citep{kenyon08}. \citet{hills88,
hills91} first predicted hypervelocity stars when
modelling the close interaction of binary star
systems with a supermassive black hole (SMBH), which
produced stars with velocities in excess of 1000 km
s$^{-1}$. \citet{yu03} proposed two additional
mechanisms: gravitational interaction between a pair
of single stars with a extreme mass ratio or stellar
acceleration by a binary SMBH system. However, they
predicted that the HVS generation rate by single
star interactions is likely low enough to be
undetectable due to the extremely small impact
parameter required. \citet{brown05} reported the
discovery of the first HVS in the Galactic Halo,
SDSS J090745.0+024507, a main sequence B star
\citep{fuentes06} which remains the fastest
observed HVS with a Heliocentric radial velocity of
831.1$\pm$5.7 \kms\ \citep{brown14}. \citet{brown14}
 assembled the most complete catalog of HVSs
with 24 confirmed objects and several additional candidates.
Several simulations \citep{brown05,Meyer14, Meyer16}
have demonstrated that it is unlikely that HVSs can
support visible bowshock nebulae due to their high
velocities inhibiting the accumulation of material
in the leading shock.

Two general classes of bowshock nebulae are generally
recognized: those supported by runaway stars, and
``in-situ'' bowshocks supported by a star overrun by an
outflow of hot gas from a star-forming or \ion{H}{2} region.
\citet{gull79} used optical emission line imaging to catalog
the first bowshock nebulae which appeared as ``distorted
interstellar bubbles''. These nebulae were observed around
the prototypical runaway $\zeta$ Oph and around the star LL
Ori situated in an outflow from $\theta^1$ Ori in the Orion
Nebula.  \citet{Povich08} cataloged six arcuate nebulae
around the star-forming regions M~17 and RCW~49  and
identified these in-situ shocks as a  distinct class. The
associated physics of these in-situ shocks are similar to
runaway bowshocks when considered within the rest frame of
the stellar source.

Using 60 $\mu$m images from the Infrared Astronomical
Satellite ($IRAS$) \citep{IRAS} \citet{vanBuren88} compiled
the first catalog of bowshock nebulae. A more complete
sample gathered from $IRAS$ images found a total of 58
bowshock nebula candidates around 188 runaway OB stars
\citep{vanBuren95}. Subsequent re-analysis of $IRAS$ all-sky
images concluded only 19 of the 58 candidate were bowshock
nebulae with 2 additional questionable candidates
\citep{Noriega97}. The increased angular resolution of
infrared surveys conducted by the Spitzer Space Telescope
($SST$) and Wide-Field Infrared Survey Explorer ($WISE$)
\citep{Wright10}, enabled the identification of small
collections of bowshock nebulae in both the LMC
\citep{Gvaramadze10} and SMC \citep{Gvaramadze11A},
generated by stars ejected from the star-forming regions NGC
6611 \citep{Gvaramadze08}, Cygnus OB2 \citep{kobul10}, NGC
6357 \citep{Gvaramadze11C}, NGC 3603 \citep{Gvaramadze13},
and Carina Nebula \citep{Sexton15}, and powered by high-mass
X-ray binaries \citep{Gvaramadze11B}, pulsars
\citep{wang13}, red \citep{Noriega97, cox12, Gvaramadze14A}
and blue \citep{Gvaramadze14B} supergiants, and the A-type
star $\delta$ Vel \citep{gaspar08}. \citet{Peri12, Peri15}
conducted the most extensive search to date for bowshock
nebulae and runaway stars by examining $WISE$ archival
images in the vicinity of several hundred runaway star
candidates identified by \citet{tetzlaff10}. The most recent
release of their E-BOSS (Extensive Bow Shock Survey) catalog
\citep{Peri15} contains a compilation of 73 bowshock nebulae
candidates found in their visual search and those previously
identified in the literature. Based on their work, it
appears that 5--10\% of runaway stars support bowshock
nebulae, though some earlier studies have suggested higher
rates of $\sim$15\% \citep{vanBuren95} or as high as
$\sim$30\% \citep{vanBuren88}.

The present paper is the first in a
series presenting and analyzing a new catalog of candidate
bowshock nebulae. Our group has taken the opposite approach
of the E-BOSS survey \citep{Peri12, Peri15} by conducting a
comprehensive visual inspection of existing space-based
infrared surveys to identify candidate bowshock nebulae and
runaway stars along the Galactic mid-Plane. In this paper,
we report the identification of 708 candidate bowshock
nebulae, 659 of which have not been previously identified in
the literature. This catalog constitutes the largest
collection of bowshock nebula candidates by an order of
magnitude. In \citet{chick16}, we present the results of
spectroscopic follow-up of nearly 100 targets, that $>$95\%
of these infrared-selected stellar sources are, in fact,
early-type massive stars, and that infrared nebular
morphology alone is enough to select massive stars with
confidence. In Section 2, we discuss the selection process
and space-based infrared datasets used for identifying
candidate nebulae. Section 3 presents the archival infrared
images of our candidates and discusses the demographics of
our sample including the sky distribution, proper motions,
and their surrounding environments. We summarize our results in Section 4.

\section{Bowshock Candidate Identification}

\subsection{Infrared Survey Data and Strategy Employed} 

Our team conducted a visual examination of mid-infrared
images from {\it Spitzer Space Telescope}
\citep[$SST$;][]{Werner04} and the Wide-field Infrared Survey
Explorer \citep[WISE;][]{Wright10} to locate bowshock nebula
candidates.  The \sst\ data included several wide-area
surveys conducted using the Infrared Array Camera
\citep[IRAC;][]{Fazio04} in its 3.6, 4.5, 5.8, and 8.0
$\mu$m bandpasses along with 24 $\mu$m data from the
Multiband Imaging Photometer for Spitzer
\citep[MIPS;][]{Rieke04}.  The \sst\ beam size at these
bands is 1.66, 1.72, 1.88,  1.98 and 6\arcsec\  FWHM, 
respectively\footnote{http://irsa.ipac.caltech.edu/data/SPITZER/docs/irac/iracinstrumenthandbook,
\\ 
http://irsa.ipac.caltech.edu/data/SPITZER/docs/mips/mipsinstrumenthandbook}.  
Survey data included images  obtained as part of the
Galactic Legacy MidPlane Survey Extraordinaire I and II
\citep[GLIMPSE, GLIMPSE~II;][]{Benjamin03} and MIPSGAL  
\citep[][]{Carey09} programs, the the $Spitzer$ Legacy
Survey of the Cygnus-X Complex \citep[][]{hora}, and the
$Spitzer$ Mapping of the Outer Galaxy
\citep[SMOG;][]{Carey08}.  Together these $SST$ surveys
cover slightly more than 304 square degrees of the Galactic
Plane  --- a large continuous strip at $-65^\circ<\ell <
65^\circ$, $| b | \lesssim 1^\circ$ (GLIMPSE and MIPSGAL), a
24 square degree regions near $\ell=79^\circ$ (Cyg-X), and a
2\degr $\times$ 10\degr\ strip centered on $\ell$=105\degr\
(SMOG).  Our search used the large mosaics at each waveband
constructed by the GLIMPSE team that are available at the
Spitzer Science Center.   The $WISE$ data include images at
the 3.4, 4.6, 12, and 22 $\mu$m bandpasses which have
beamsizes of 6.1\arcsec, 6.4\arcsec, 6.5\arcsec, and
12.0\arcsec\ FWHM, respectively.  Our search of the $WISE$
all-sky survey images (about 460 square degrees of the Outer
Galaxy) used 1\degr\ mosaics in regions along the Galactic
Plane not covered by Spitzer and the Centre de Donn\'ees
astronomiques de Strasbourg's 
Aladin\footnote{http://aladin.u-strasbg.fr/}  interface
which allows a viewer to rapidly browse large regions of the
sky.  Given the concentration of  bowshock nebulae very near
the Galactic Plane (Section \ref{loc.sec}),  our search of
the $WISE$ data concentrated on regions $| b |$ $<$ 2\degr,
but we also performed a cursory search using the Aladin
\citep{Bonnarel00} Lite interface up to $| b |$ $<$
10\degr.  Only a few objects were discovered 
beyond $b=$ $\pm$1\degr.   

We searched the $SST$ mosaics for 1) arc-shaped or
lima-bean-shaped nebulae at 24 $\mu$m  and/or 8.0 $\mu$m
that had 2) a high degree of symmetry with 3) at least one
prominent star located near the axis of symmetry visible at
4.5 $\mu$m.  For the majority of candidates, there is a single
bright star located along the axis of symmetry;  it
constitutes the obvious early-type stellar candidate, even
though there is no spectral information available for the
vast majority of candidates.  In some cases there are
multiple candidate stars within the arc, and in a few cases
there are no prominent stellar candidates, especially in the
inner Galaxy where the stellar density is high and even
early type stars may  appear faint because they are located
at large distances. 

Our initial visual search team interpreted these three
criteria liberally.  The Galactic Plane exhibits very
complex structures at 8.0 and 24 $\mu$m that vary greatly in
surface brightness over small angular scales, producing a
large number of initial candidates that do not meet one of
more of the three search criteria.  The final decision
regarding which objects met all three criteria was
ultimately  made by the senior author (HAK).  Relaxing these
criteria would substantially increase the candidate list by
at least a factor of two.  The search process involved
examining color representations of the 4.5, 8.0, and 24
$\mu$m images displayed with a square-root color stretch. 
Owing to the high dynamic range  of the mosaics, it was
necessary to use a variable maximum and minimum intensity
levels to achieve the needed contrast, depending on the
surface brightness in each region.   Often in regions of
large dynamic range several different scalings were used. 
This was necessary to find nebulae superimposed on regions
with high background levels (e.g., within \ion{H}{2}
regions).  No quantitative thresholds were used in the search
process; indeed, the rms noise levels are not well
defined, varying significantly  on small spatial scales
where regions of very high and very low surface brightness
appear adjacent to one another.  The maximum angular size of
a detectable candidate is limited by the size of the survey
mosaics at about 2 degrees; in practice, there are no
candidates larger than 8.6\arcmin.   The angular resolution
of the $SST$/$WISE$ surveys, described above, sets the lower
limit on the angular size of detectable candidates. 

 Two or more team members searched each region of
the Plane. In comparing candidate lists identified by each
searcher we concluded that there was at least a 50--75\%
overlap between the candidate lists.  Each searcher
identified candidates not found by the others.  We estimate
that use of additional searchers could increase the current
candidate list, most of this gain coming in areas of the
Plane dominated by high-surface brightness \HII\ regions.  A
search such as this can never be deemed ``complete'' given
the inherent subjectivity of what constitutes ``arc-shaped''
or ``symmetric'', especially in regions having large changes
in surface  brightness over small angular scales  where
genuine bowshock nebulae may be confused with unrelated
foreground or background emission.

All $SST$ objects were discovered initially in the 24 $\mu$m
bandpasses where emission is either thermal in nature  or
from (semi-)stochastic heating of small dust grains.  Only a
minority of the bowshock candidates exhibit morphologically
similar counterparts in the 8.0 $\mu$m band, which
predominantly includes a broad solid-state emission feature
from polycyclic aromatic hydrocarbons (PAHs).  However, the
8.0 $\mu$m band may also include a contribution from very
hot dust near the sublimation temperature.    

Our search of the WISE images concentrated on the 22 $\mu$m
(W4) band for nebula identification and the 4.6 $\mu$m (W2)
band for identification of a stellar source.   We find that
nearly all candidates identified in the W4 band also had
counterparts in the 12 $\mu$m (W3) band, which may include
both PAH and hot dust emission.  Owing to the superior
angular resolution of $SST$ compared to WISE -- 6.0\arcsec\
at 24 $\mu$m versus 12.0\arcsec\ FWHM at 22 $\mu$m -- we
gave preference to the $SST$ data (where available) and did
not search the overlapping $WISE$ images.  

Because the $WISE$ beamsize is twice that of the $SST$ at
the longest wavelengths, objects that would appear arcuate
in $SST$ imaging might appear circular in $WISE$ and,
therefore, be rejected from the catalog.  This leads to a
kind of systematic decrease in the completeness (at fixed
angular sizes of 6--12\arcsec) for the regions surveyed only
with $WISE$.  Similarly, we find a large number of small
circular  24 $\mu$m nebulae near the angular resolution
limit of $SST$ that might appear arcuate if
viewed under higher angular resolution.

Additionally, bowshock candidates listed in the literature
were examined  and retained if they met our morphological
selection criteria described above. Our survey of the
literature included the compilations of
\citet[][Pe15]{Peri15}, \citet[][S15]{Sexton15},
\citet[][Pe12]{Peri12}, \citet[][G11]{Gvaramadze11C}, 
\citet[][K10]{kobul10},  \citet[][G08]{Gvaramadze08},
\citet[][P08]{Povich08}, \citet[][CP07]{Comeron07},
\citet[][N97]{Noriega97}, and \citet{vanBuren88}.  The final
collection tabulated here does not include many of the
candidates listed in these works, primarily because they
fail to meet one of more of the morphological criteria.  

\subsection{Search Results}

Table~\ref{bigtable.tab} lists parameters for  the probable
central stars of the \numtot\ candidate bowshock nebula. 
The first twenty rows appear in the printed Journal to give
guidance as to the table format and content.  The entire
table is available in digital format in the electronic
edition.   Column 1 is a sequential identification number
unique to this work, column 2 is a generic name in Galactic
coordinates, while columns 3 and 4 list the J2000 right
ascension and declination  for the putative central star.  Column 5 gives the publication 
where the candidate is first identified as a bowshock
candidate.   A ``T'' indicates ``this work'', and other
abbreviations refer to published works, as referenced fully
in the table footnotes.  Of the \numtot\ tabulated objects,
\numnewhere\ are identified herein as bowshock candidates
for the first time.  Column 6 lists another common alias for
the central stellar source, if one exists in the literature,
such as an HD (Henry Draper Catalog) designation.  Fully
\numnoid\ of the \numtot\ stellar sources have no alternate
designation in the literature, indicating that the vast 
majority of the objects are faint and highly reddened. 
Column 7 is a single character designating whether the
nebular object has an $SST$ IRAC 8.0 counterpart with a
similar morphology (Y),  has no detection above background
levels at this waveband (N), or has no data from IRAC
(-).\footnote{Essentially all of the objects detected in the
WISE 22 $\mu$m band also have counterparts in the WISE 12
$\mu$m band.  Therefore, we do not tabulate detections for
the WISE bandpasses.}  We find that \numeight\ of the
\numtotirac\ objects having IRAC data show a nebula at 8.0
$\mu$m, indicating a probable PAH contribution.  Column 8
contains a single alpha-numeric code, ``C'', in cases where
the probable central stellar source is uncertain owing to
faintness or to the presence of multiple stars of similar
magnitude near the nebular axis; \numC\ objects fall into
this category.  In such cases we adopt the brightest point
source nearest the axis of symmetry as the most probable
star, but we urge caution with regard to the certainty of
this identification.  Columns 9 and 10 list the distance
$R_0$ (in arcsec) from the putative central star to the
apsis of the candidate bowshock and the position angle (in
degrees from N toward E)  of this vector in equatorial
coordinates.  These values were measured by eye and carry
typical uncertainties of 1\farcs5 and 5\degr, respectively. 
Columns 11 and 12 list the 2MASS H-band and IRAC 4.5/WISE
4.6 $\mu$m magnitudes of the central star, obtained from the
2MASS \citep{2MASS} and GLIMPSE/WISE catalogs,
respectively.  In just a handful of cases (\numNOH), no
H-band source is measured, implying an extremely red, highly
extinguished object.  Column 13 lists an estimate of the K-band
extinction to each source, inferred using the H$-$4.5 color
and the Rayleigh-Jean Color Excess formulation  of
\citet{Majewski11}.  

The final column of Table~\ref{bigtable.tab} provides
additional descriptors that characterize the local
environment of each stellar candidate.  Column 14 contains
an alphanumeric code that describes the immediate
environment of each candidate as either isolated (I; \numI\
instances), directly facing a large \ion{H}{2} region within
about 10 arcminutes (FH; \numFH\ instances), facing a
bright-rimmed cloud prominent at 8.0 $\mu$m within several 
arcminutes (FB; \numFB\ instances), or situated within a
giant \ion{H}{2} region (H; \numH\ instances).  In some
cases two designations could apply, for example, a candidate
that lies within an \ion{H}{2} region that also faces a
bright-rimmed cloud.  In such cases FB is given preference
over H.   Objects designated FH may be instances of the
``in-situ'' bowshocks \citep[][]{Povich08}.  These
descriptors are qualitative judgments of the candidates'
immediate environments intended to elucidate possible
physical origins for what we regard as potentially different
classes of bowshock nebulae.  Given the unknown distances of
each object and complex sightlines in the Galactic Plane,
physical connections between bowshock candidates and other
objects nearby on the sky are necessarily uncertain. 
Section \ref{env.sec} will discuss example objects in each
of these five categories in more detail.  

\section{Properties of the Bowshock Candidates}

\subsection{Infrared Images} 
\begin{figure}  
\plotone{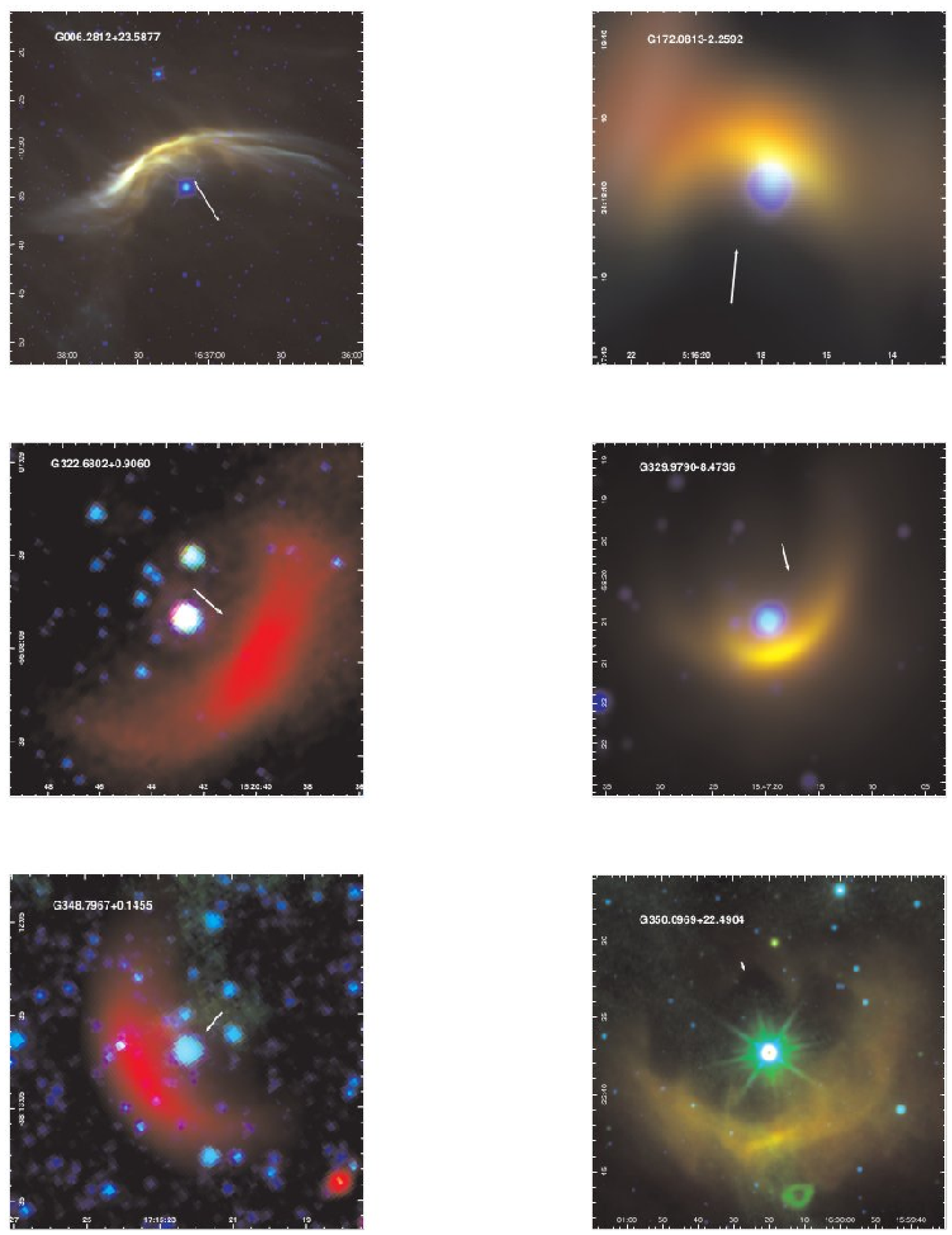}  
\caption{Six
prototype bowshock nebulae, with the  colors representing
either the 24 $\mu$m \sst\ or 22 $\mu$m $WISE$ band in red,
either the 8 $\mu$m \sst\ or 12 $\mu$m  $WISE$ image in
green, and the 4.5 $\mu$m \sst/$WISE$ image in blue.  These
six objects are  $\zeta$ Oph (G006.2812+23.5877; upper
left),  AE Aur (G172.0813$-$02.2592; upper right),  HD136003
(G322.6802+00.9060; center left), HD150898
(G329.9790$-$08.4736; center right), HD155755
(G348.7967+00.1455; lower left), HD143275
(G350.0969+22.4904; lower right).  Arrows in some panels
indicate the direction of proper motion, if known.  
\label{prototypes} }   
\end{figure}

Figure~\ref{prototypes} shows a
collage of three-color representations for six objects from
Table~\ref{bigtable.tab} that may be regarded as
prototypical bowshock nebulae.  The angular scale of each
panel differs to accommodate the size of each object.   Each
panel displays a highly resolved, unmistakable arc-shaped
nebulae and a bright central star with a substantial proper
motion consistent with the nebular bowshock orientation.  
These six objects are  $\zeta$ Oph (G006.2812+23.5877; upper
left),  AE Aur (G172.0813$-$02.2592; upper right),  HD136003
(G322.6802+00.9060; center left), HD150898
(G329.9790$-$08.4736; center right), HD155755
(G348.7967+00.1455; lower left), HD143275
(G350.0969+22.4904; lower right ), as ``prototypical''
examples of bowshock nebulae drawn from the literature.  We
use a color scale that depicts the $SST$ 24 $\mu$m or $WISE$
22 $\mu$m image in red, the $SST$ 8.0 or $WISE$ 12 $\mu$m
image in green, and the $SST$ 3.6 or $WISE$ 3.6 $\mu$m image
in blue. A white arrow depicts the proper motion vector
magnitude and direction.   Images of G322.6802+00.9060 and
G348.7967+00.1455 use IRAC data so the nebula is dominated
by 24 $\mu$m emission (red) and appears red, while the other
four  use $WISE$ data, and the combination of 22 $\mu$m
(red) and 12 $\mu$m (green)  emission results in a yellowish
appearance.   

Figure~\ref{prototypes} reveals that only a minority of the
objects display the ideal (``theoretical'') bowshock
morphology ---  e.g., see Figure~1 of \citet{Wilkin96} or
Figure~9 of  \citet{Meyer16} or Figure 3 of
\citet{Acreman16}  for a variety of viewing inclination
angles). The $\zeta$ Oph nebula (upper left) appears
flocculent and significantly asymmetric, perhaps because it
is highly resolved as the nearest of the bowshocks at a
distance of only 140 pc, based on its Hipparcos parallax
\citep{Perryman97}.   Most of the nebulae are, to some
degree, asymmetric. In some cases, the star appears slightly
displaced from the axis of symmetry.   \citet{Wilkin00}
discusses such a displacement as the natural result of
either a density gradient in the ambient medium transverse
to the star's motion or a stellar wind that is
anisotropic.  

A digital Appendix contains images of all \numtot\ candidate
objects from Table~\ref{bigtable.tab} in Figures~A.1 through
A.118  using the same  color scheme as
Figure~\ref{prototypes}.   These figures depict an angular
field of view of 2 arcmin square in most cases to facilitate
easy angular size comparison, but for some objects a window
of 4\arcmin\ or 8\arcmin\ is necessary to include the entire
nebula or pertinent features in the immediate environment. 
For a few objects observed by the $SST$ IRAC  but not at
MIPS  24 $\mu$m, we use the $WISE$ 22 $\mu$m image instead.
In a just a few instances, the images are constructed using
the IRAC 8.0/4.5/3.6 $\mu$m images. In some panels a white
arrow depicts the proper motion vector magnitude and
direction, if proper motion data are available and the
measured proper motion is larger than the uncertainties.   
The vast majority of the \numtot\ candidate bowshock
nebulae  have small angular sizes and are, consequently,
less resolved than the prototypical examples;  nevertheless,
they share the overall shape, colors, and morphological
symmetry of the prototypes.  

\subsection{Candidate Bowshock Distribution \label{loc.sec}}

\begin{figure}  
\gridline{\fig{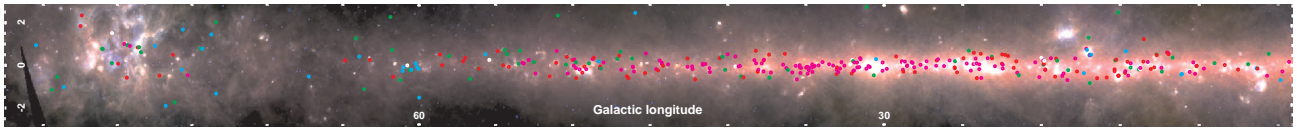}{1\textwidth}{(a)} }
\gridline{\fig{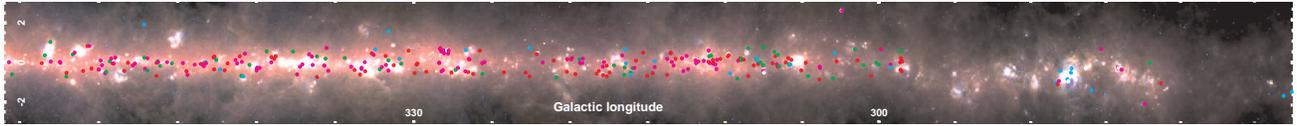}{1\textwidth}{(b)} }
\caption{Candidate bowshock nebula in the inner two Galactic quadrants 
overlaid on
a color representation of the $IRAS$ 25/60/100 $\mu$m maps
in blue/green/red.   Colors denote the
extiction values estimated from the $H-$4.5 $\mu$m color
excess: cyan--\av$<$2; green--$2<$\av$<5$;
red--$5<$\av$<10$; magenta--\av$>10$; white--no data. 
\label{allcolor180} }  
\end{figure}

Figure~\ref{allcolor180} plots the positions of the
candidate bowshock nebulae on the Galactic Plane for the
\numInnerG\ objects that lie in the inner two Galactic
quadrants at longitudes $-90^\circ<\ell<90^\circ$.   Colors
denote the extiction values estimated from the H$-$4.5
$\mu$m color excess: cyan--\av$<$2; green--$2<$\av$<5$;
red--$5<$\av$<10$; magenta--\av$>10$; white--no data.  The
points appear superimposed on a three-color representation
of the Galactic Plane with the Infrared Astronomical
Satellite  ($IRAS$) maps at 25/60/100 $\mu$m in
blue/green/red, respectively.  The distribution of objects
is non-uniform, even allowing for the fact that  they were
selected from surveys covering only a portion of the Plane,
principally within one degree of $b$=0\degr\  and within
60\degr\ of Galactic Center.  For  example, there is an
overdensity  of objects in the vicinity spiral arm
tangencies near $\ell$=25--30\degr\ where the path length
through spiral arms is long.   The Cygnus-X region near
$\ell$=79\degr\ also exhibits an abundance of objects,
probably the result of having multiple nearby (1.2--1.5 kpc)
regions of massive star formation.   The Carina star forming
region near $\ell$=287\degr\ also exhibits a large number of
candidates.  A close examination of 
Figure~\ref{allcolor180}  further reveals that objects
having similar  extinctions cluster together on the sky, a
probable signature  of clusterings of massive stars at
similar distances and reddennings.  

\begin{figure}  
\plotone{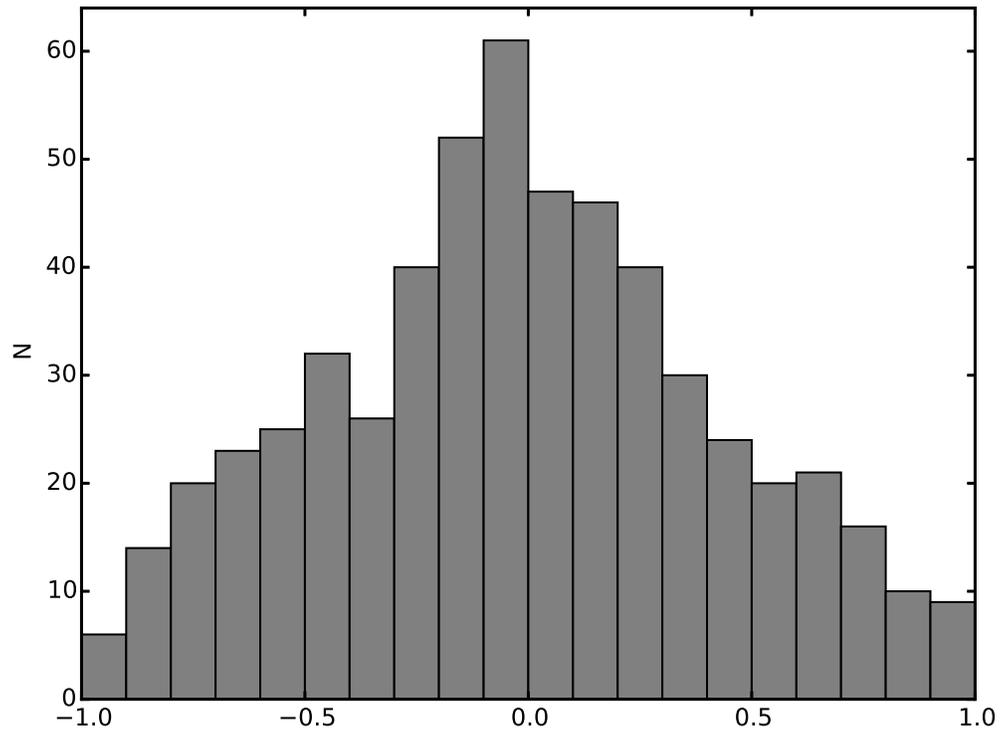}  
\caption{Histogram of Galactic latitudes for bowshock
candidates within the footprint of the
GLIMPSE and MIPSGAL surveys: $-60^\circ<\ell<60^\circ$,$-1^\circ<b<1^\circ$.
\label{histB} }  
\end{figure}

Figure~\ref{histB} is a histogram of the Galactic latitudes
for the 561 candidates that lie within the footprint of the
GLIMPSE and MIPSGAL surveys where coverage is complete: 
 $-60^\circ<\ell<60^\circ$,$-1^\circ<b<1^\circ$. The
histogram is strongly peaked near $b$=0\degr,  indicating that
very few objects meeting our selection criteria are likely
to be found at larger latitudes.  The angular scale height 
for these objects is approximately 0.6\degr\ from mid-Plane,
identical to the scale height for interstellar bubbles
and \hii\ regions \citep{Churchwell06}.  

\subsection{Bowshock Orientations}

\begin{figure}
\gridline{\fig{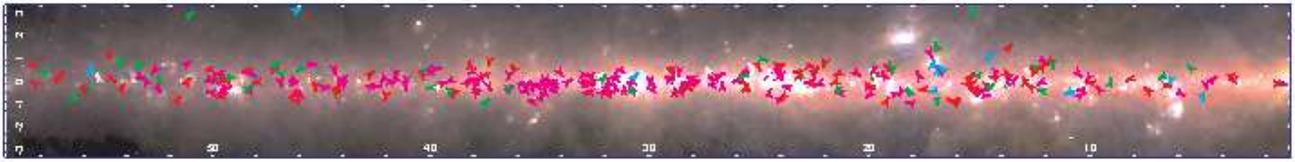}{1.\textwidth}{(a) } }
\gridline{\fig{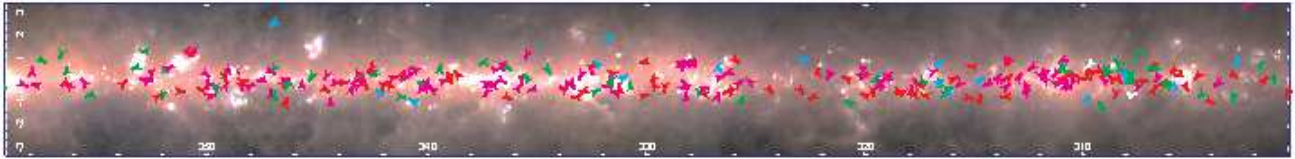}{1.\textwidth}{(b)} }
\caption{ Locations of bowshock candidates drawn as arrows to indicate 
the morphological orientation of the
nebula.  Colors designate interstellar extinction values, 
as in Figure~\ref{allcolor180}.  The upper panel shows
 a portion of the Plane at 
$0^\circ<\ell<60^\circ$ while the lower panel shows the 
$300^\circ<\ell<360^\circ$
region.   \label{PAmap} }
\end{figure}

Figure~\ref{PAmap} displays the positions of bowshocks
candidates at $0^\circ<\ell<60^\circ$ ({\it upper panel})
and  $300^\circ<\ell<360^\circ$  ({\it lower panel}). 
Symbol color and the  background three-color IRAS image is
the same as  in Figure~\ref{allcolor180}.  Vector directions
indicate the  orientation of the bowshock nebulae as
determined by the IR morphology.   This figure further
illustrates that objects with similar extinctions often
cluster together on the sky.  Furthermore, there are also
clusters of objects that have similar orientations.   One
notable example of this is at the base of the giant gas
pillar in M~16 near $\ell$=16.8\degr, $b$=0.6\degr\ where
four objects all point toward the source of radiation that
has sculpted  the pillar.     

\begin{figure}  
\plotone{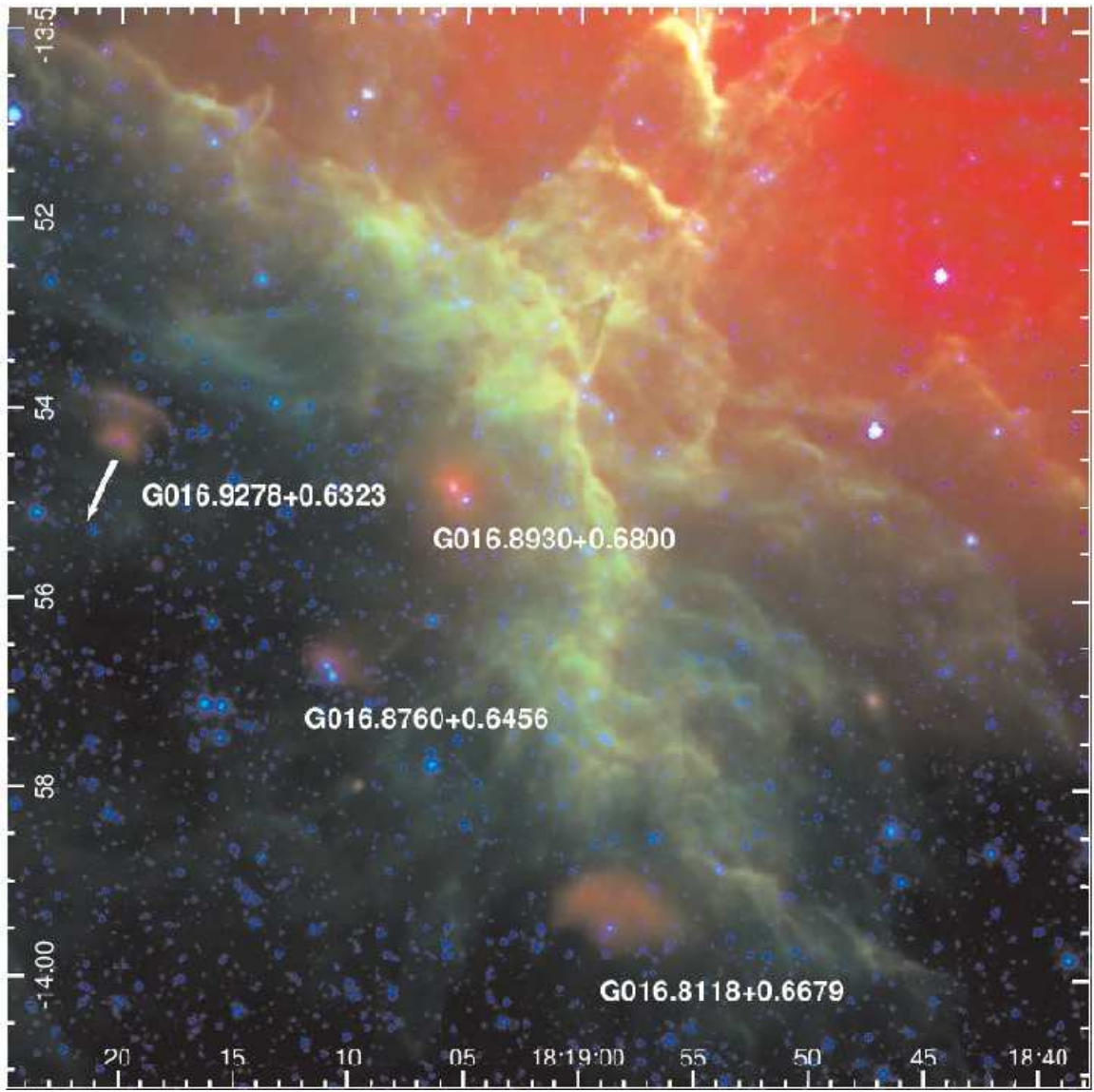}  
\caption{Three color depiction of four candidate bowshock nebulae
exhibiting a similar orientation 
at the base of the giant pillar in M~16.  Red/green/blue represent the
\sst\ 4.5, 8.0, and 24 $\mu$m band images.  The arrow indicates the
proper motion of the one star having measured values. 
\label{M16four} }  
\end{figure}

Figure~\ref{M16four} depicts four candidate bowshock
nebulae  (G016.8118+00.6679, G016.8760+00.6456,
G016.8930+00.6800, and  G016.9278+00.6323) in the same
field, all pointing in the same direction as the pillars.   
The central stars of the bowshocks are HD~168183
(O9.5III+B4V), BD$-$13~4936 (B0e), BD$-$13~4934 (B1Vp), and
BD$-$13~4937 (B1.5V), respectively, all early-type stars.
BD$-$13~4937 has a measured proper motion directly {\it
away} from the source of radiation in this young star
forming region.  These are excellent candidates for
``in-situ'' bowshocks  \citep{Povich08} formed by the
interaction between the star's own wind and an outflow from
the M~16 \hii\ region.   In regions like this one, bowshocks
exhibit a similar orientation  not because they share a
common space velocity but because a common source of
external momentum flux impinges upon them.  There are many
other possible ``in-situ'' bowshocks in our sample, and
these are flagged by FH in Table~\ref{bigtable.tab}.   

\begin{figure}  
\plotone{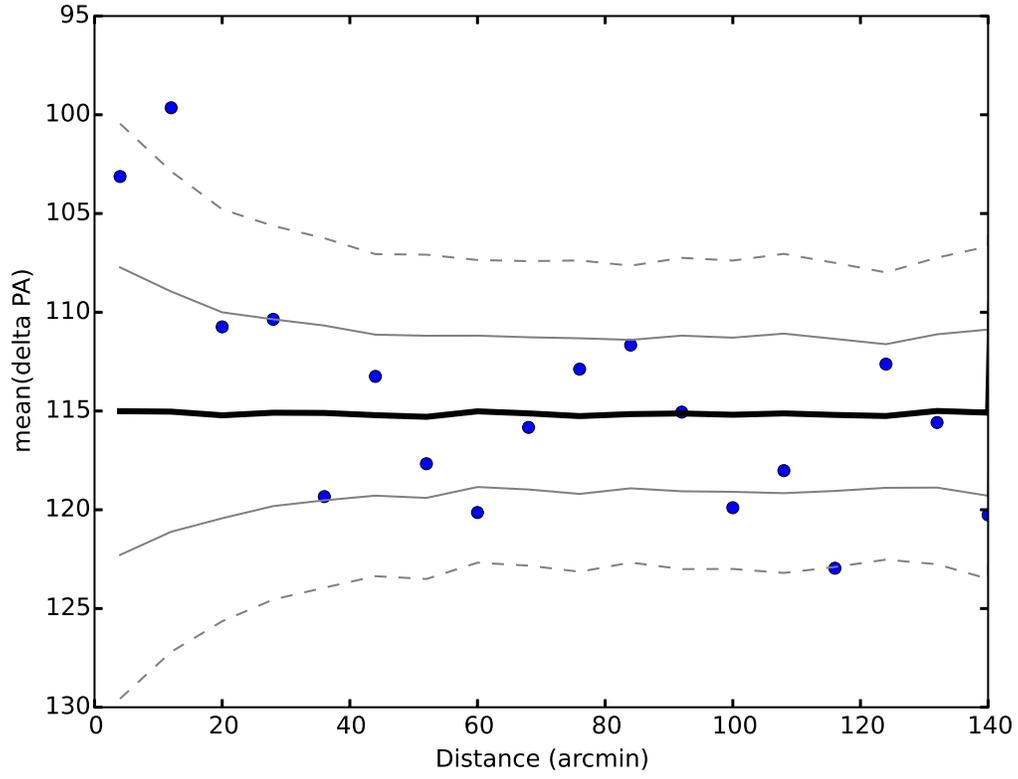}  
\caption{Mean deviation in bowshock position angle from nearest neighbors 
as a function of angular separation in arcminutes.  Points show
the data while the heavy black line shows the mean of 1000 random iterations
simulating the null hypothesis that the position angles are uncorrelated.  
The gray and dashed lines illustrate the 1$\sigma$ and 2$\sigma$
deviations from the mean.  
\label{structure} }  
\end{figure}

In order to test the hypothesis that bowshock  orientations
in this field and others  may be correlated on small
angular scales, we computed the difference in position angle
between each pair of the N=\numtot\ nebulae
(N$\times$(N$-$1)/2 unique pairings)  and then plotted the
mean $\Delta$PA versus  angular separation in 8\arcmin\ bins
in Figure~\ref{structure}.  For comparison to the null
hypothesis (that the position angles are uncorrelated) we
randomly shuffled the position angles among the objects 1000
times and computed the mean $\Delta$PA  and dispersion in
each bin.  Figure~\ref{structure} plots the data (blue
points) and the results of the random Monte Carlo 
iterations, where the heavy black line depicts the mean
difference in position angle. The  thin solid and dashed
lines depict the 1$\sigma$ and 2$\sigma$ deviations about
the mean.  At small angular separations $<$20\arcmin\  the
two blue points lie near the  2$\sigma$ envelope, while at
larger angular separations the points scatter about the mean
expected in the case of random orientations.   This
demonstrates that bowshock positions angles are correlated
on small angular scales, as would be expected if some are
produced by large-scale external influences such as outflows
from nearby \hii\ regions.  

\begin{figure}  
\plotone{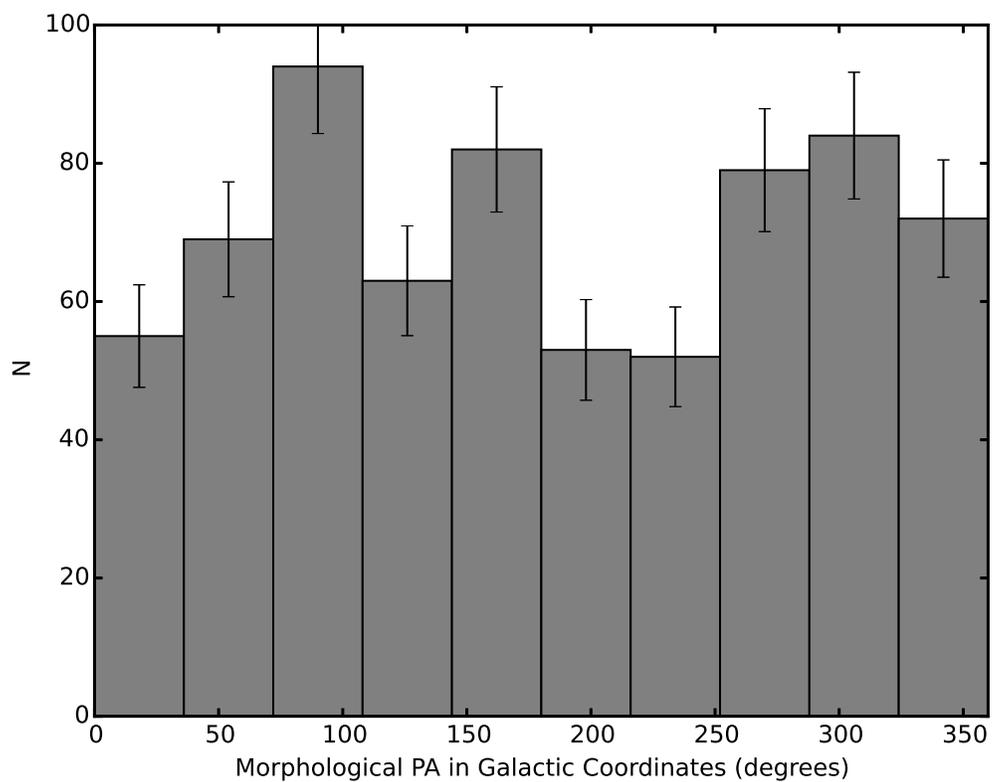}  
\caption{Histogram of orientations for candidate bowshock nebulae in Galactic
coordinates. Error bars indicate the Poisson uncertainties in each bin. 
\label{histPA} }  
\end{figure}

We also investigated whether the bowshock nebulae display
preferred orientations in the Galactic coordinate system.
Figure~\ref{histPA} displays a histogram of the
morphological position angles, in Galactic  coordinates, for
all \numtot\ objects.  Error bars designate Poisson
uncertainties for each bin.  While the overall distribution
is generally consistent with random, there is a small excess
of objects at 90\degr\ and 270\degr\ orientations, parallel
to the Plane.  

\subsection{Bowshock Sizes}

\begin{figure}  
\plotone{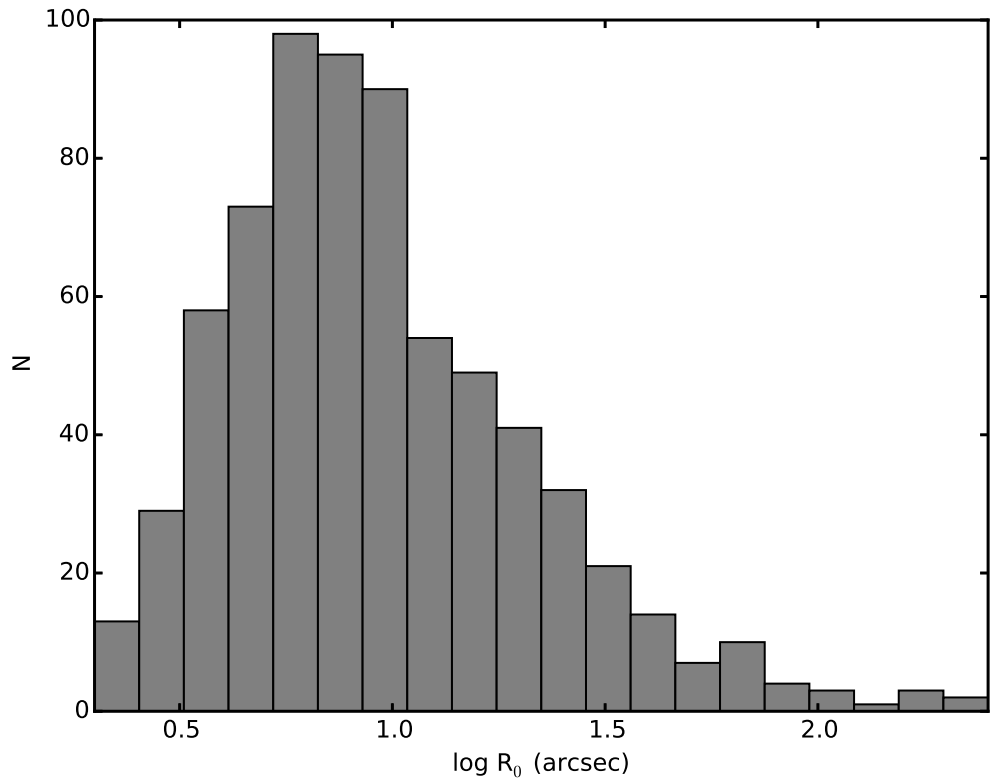}  
\caption{Histogram of bowshock sizes, as indicated
by the standoff distance, $R_0$, in arcseconds between
the central star and apsis of the nebular arc.
\label{sizes} }  
\end{figure}

Figure~\ref{sizes} shows a histogram of the base 10
logarithm of the standoff distance, $\log R_0$, in
arcseconds, describing the separation of the central star
from the  apsis of the arc-shaped nebula.  The peak of the
distribution  in the range 5--12\arcsec\ indicates that the
vast majority of the objects are compact, sufficiently small
that the angular resolution of \sst\ is required to discern
the arc-like shape of the nebula.   The distribution
declines smoothly from 12\arcsec\ to about 50\arcsec, 
beyond which a small number of very large arcs with sizes up
to 500\arcsec\ populate the distribution more or less
uniformly.  On the smaller end of the distribution the
number of objects drops steeply from sizes of 5\arcsec\ to the
resolution limit if the IRAC instrument at about 2\arcsec.   

\begin{figure}  
\plotone{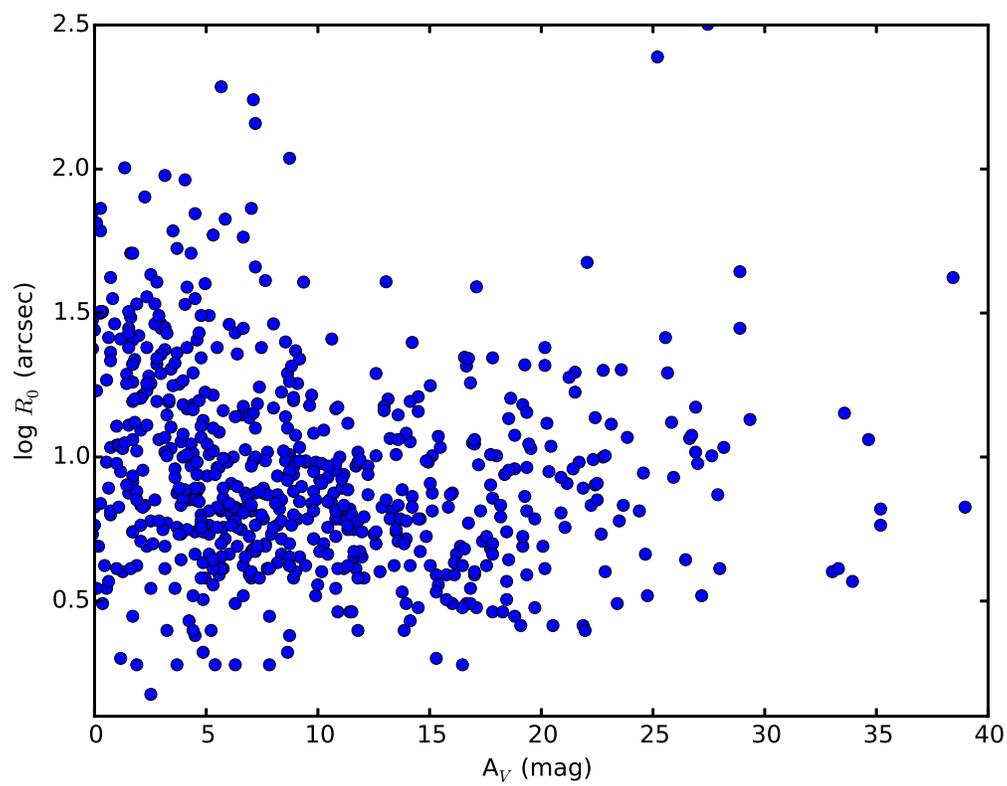}  
\caption{Base 10 log of bowshock size versus inferred extinction, \av.
\label{avr0} }  
\end{figure}

Figure~\ref{avr0} plots the angular size of the objects,
$R_0$, versus  extinction toward the star, \av, as inferred from the $H-$4.5
$\mu$m color excess, assuming a standard extinction law with
\av = 9$\times$$A_{\rm K}$.  A weak correlation is evident, with
the largest nebulae being the least extinguished and the
smallest nebulae having  preferentially larger extinctions. 
If extinction of the starlight were a proxy for distance, we might expect
that the most extinguished objects would also have the smallest
angular sizes, but no strong trends are evident.

\subsection{Bowshock Environments \label{env.sec}}

\begin{figure}  
\plotone{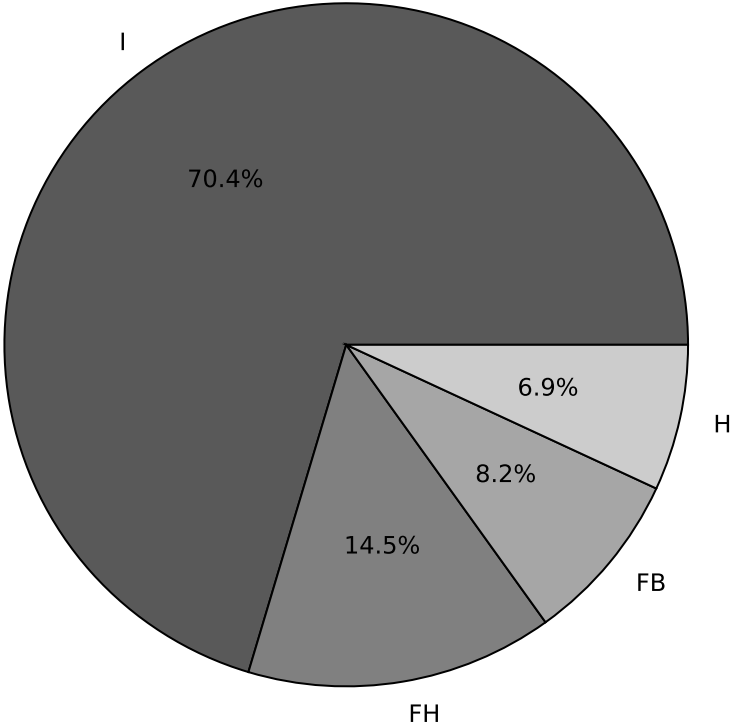}  
\caption{Fraction of candidates located in apparently isolated environments (I), facing \hii\ regions (FH),
facing bright-rimmed clouds (FB), and located within \hii\
regions (H).  FB frequently entails H.  
\label{environ} }  
\end{figure}

Figure~\ref{environ} depicts the fraction of the bowshock
candidate sample that lies in apparently isolated
environments (I; 70.4\%),
facing \hii\ regions (FH; 14.5\%), facing bright-rimmed
clouds (FB; 8.2\%), and located within \hii\ regions (H;
6.9\%).  Many objects designated as FB also lie within \hii\
regions, so FB may entail H; FB and H together make up
15.3\% of the sample. Isolated objects (\numI) make up more
than two-thirds  of the sample.  However, this designation
does not necessarily mean that the object is  physically
isolated from other potential astrophysical influences. In
the complex regions near the Galactic Plane, most candidates
are seen in projection near dark clouds, star forming
regions, and other prominent features.  Rather, the isolated
designation simply means that the object does not obviously
appear to lie within or to directly face a visible \hii\
region, star or bright-rimmed cloud.  As such, these are the
best candidates for being genuine runaway stars with space
velocities sufficient to sweep up ambient material into an
infrared-emitting bowshock  nebula.

About 14.6\% (103 objects) of the sample exhibits a
bowshock orientation that faces a prominent \hii\ region
within $\simeq$10\arcmin. In such close proximity,  outflows
from a hot ionized bubble  could plausibly interact with the
wind of the OB star, producing an ``in-situ''  type bowshock
even if the star does not have a high space velocity.  The
association  of a bowshock nebulae with a particular  \hii\
region is necessarily speculative, but the relative
orientation occurs frequently enough in our sample that the
physical nature of this class of objects is highly
suggestive.

\begin{figure}  
\plotone{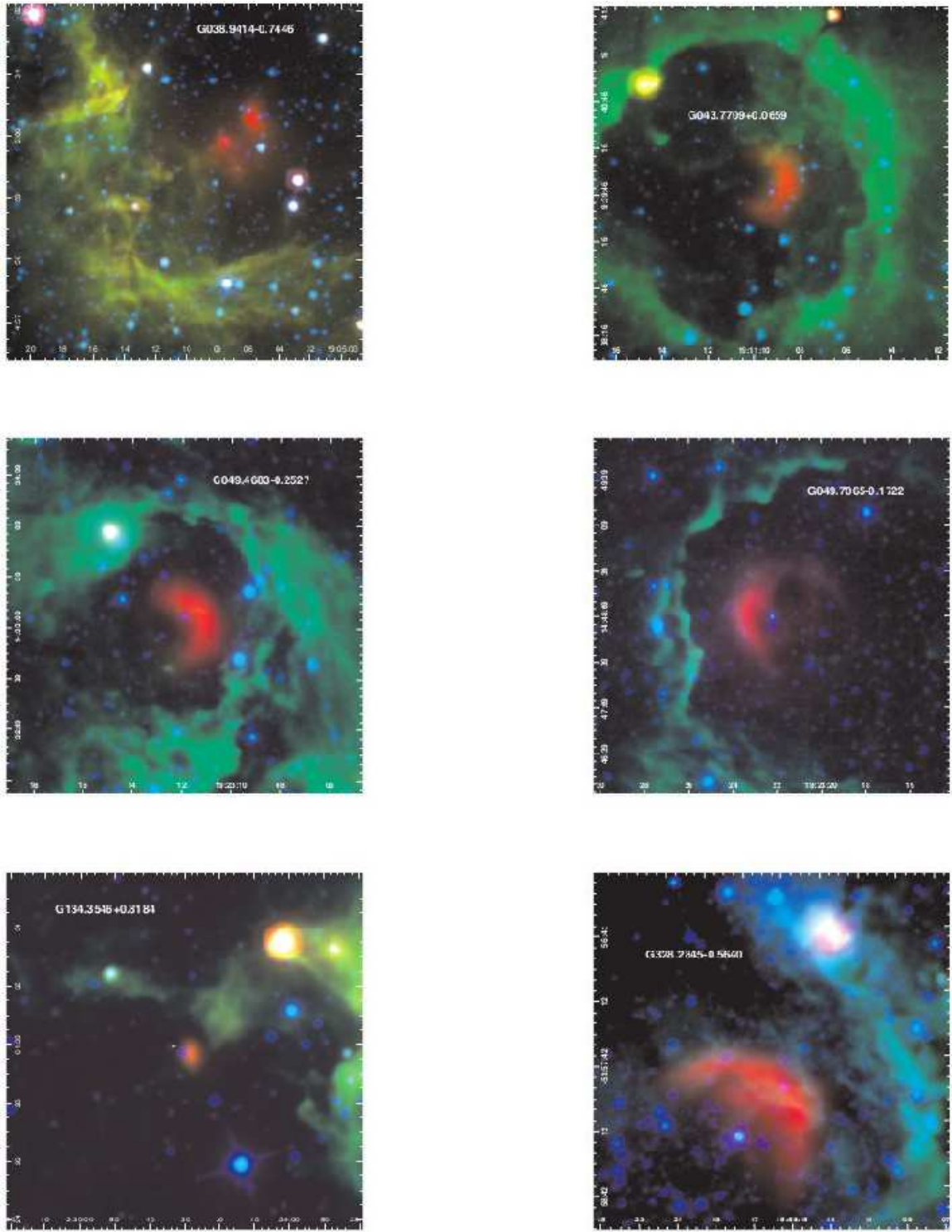}  
\caption{Examples of six candidate bowshock nebulae that face an
8 $\mu$m bright-rimmed cloud where the shock may be cause by the stellar wind interacting with a
photoevaporative flow from the molecular cloud interface.  
These may be considered a subclass of in-situ bowshocks. 
The color scheme follows Figure~\ref{prototypes}.  
\label{examples.FB} }  
\end{figure}

Eight percent (58 objects) of the sample faces a
bright-rimmed  cloud within several arcminutes.
Bright-rimmed features  represent the photo-excited surfaces
of molecular clouds emitting strongly in the 8.0 $\mu$m
bandpass by virtue of the solid state features of PAHs. They occur commonly at the
peripheries of \hii\ regions where massive stars create
cavities or bubbles in the surrounding gas, e.g, the
collection of interstellar bubbles cataloged by
\citet{Churchwell06}.   Figure~\ref{examples.FB} shows six
objects that face bright-rimmed clouds, with the color
scheme as in Figure~\ref{prototypes}.   Frequently, this
class of objects is located {\it within} an \hii\ region.  
Physically, this class of objects may arise when an OB
star's wind interacts with a photo-ablated flow from the
molecular cloud boundary produced by the UV photons from the
same star.  The phenomenon may be similar to the stationary
shocks visible as ionized arcs  at the interface between the
wind from $\theta^1$ Ori C and the proplyds in the Orion
Nebula \citep{Bally98, Garcia01}.  Figure~3 of
\citet{Garcia01} conveniently illustrates the location of
one object, proplyd 180-331, that hosts an arc facing the
ionization front in the OMC1 molecular cloud rather than the
facing the star $\theta^1$ Ori C.  Given that the phenomena
operating in our subsample of IR-selected bowshock may be
similar, we term this subclass of of in-situ bowshock nebulae ``photo-evaporative
flow'' (PEF) bowshocks.  However, in Orion, the arcs are
very thin features visible primarily in $H\alpha$ and
\ion{O}{3}, whereas our sample is prominent in the mid-IR
dust continuum and the ionization structure is not presently
known.  One prediction following from this interpretation is
that the PEF bowshock candidates, designated here as FB, do
not have high space velocities relative to their local
environment.    

The remaining 7.0\% (49 objects) appear to lie within \hii\
regions or interstellar bubbles, as outlined by either
bright-rimmed clouds or by pillars of photo-excited gas.
While they do not obviously face a bright-rimmed cloud
at the periphery of the \hii\ region, they may
be produced where a stellar wind interacts with a
high-velocity flow of material, including champagne flows. 
Even so, these objects may overlap with the FB
subsample which are frequently found within \hii\ regions.

\subsection{Proper Motions}

\begin{figure}  
\plotone{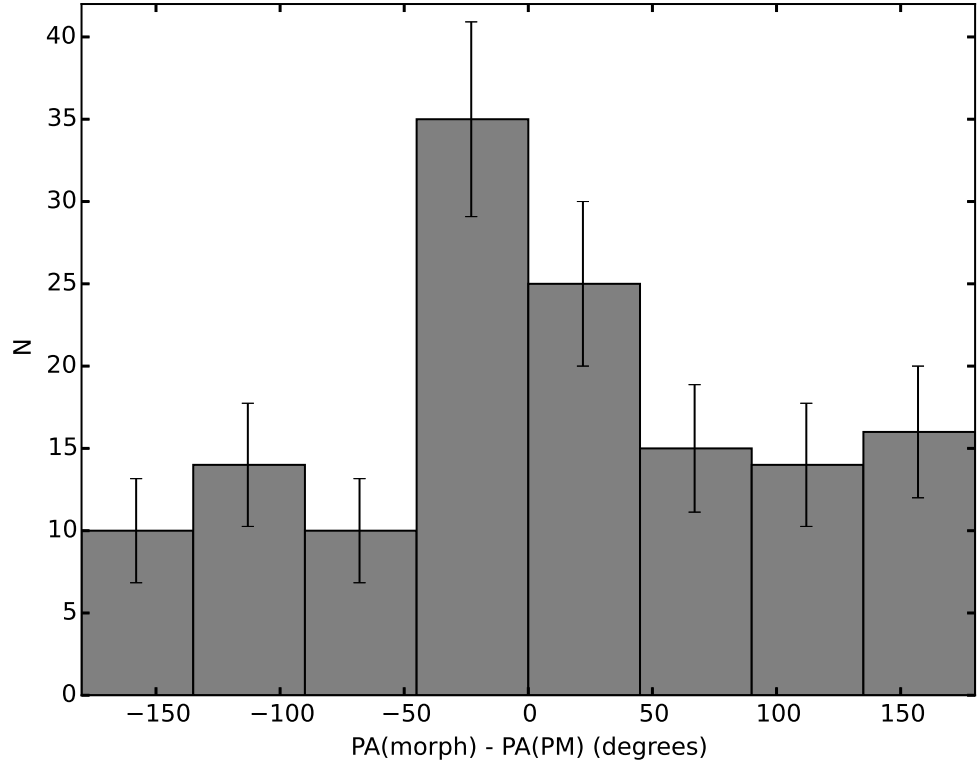}  
\caption{Difference between the mid-IR morphological axis and proper motion
vector of the central star for 139 candidate bowshocks having proper motion
position angles with uncertainties less than 45\degr.  
The statistically significant excess near zero degrees constitutes evidence
that some fraction of these objects are runaway stars having high space
velocities in a direction indicated by the IR morphology.     
\label{deltaPA} }  
\end{figure}

Following the analysis of \citet{vanBuren95}, we  collected
proper motion measurements for the stellar  counterparts to
each bowshock nebula.  In this analysis we use proper motion
data from the Hipparcos \citep{Perryman97}, UCAC4
\citep{Zacharias13}, and the USNO B2 \citep{Monet03}
catalogs, in order of preference.    Of the \numtot\
candidates, only 40\% (\numPM) have proper motion data of
any kind, and of these only 139 have sufficiently small
uncertainties that the  proper motion position angles are
known to better than 45\degr.   Nevertheless, we computed
the measured the angle between the nominal  orientation
angle of the putative bowshock nebula (column 10 of 
Table~\ref{bigtable.tab}) and the proper motion vector.  
Figure~\ref{deltaPA} displays a histogram of this
distribution in 45\degr\ bins between $-$180\degr\ and
180\degr.  Error bars depict the uncertainties, taken to be
the square root of the number of objects in each bin.   A 
clear excess in the two bins on either side of zero
indicates a preference for proper motions to coincide with
the expected direction inferred from the infrared
morphology.   This suggests that at least 25\% of the
objects are good candidates for being runaway stars with
high space velocities in the direction of apsis of the
infrared arc.  However, the fraction of sources in this plot
having relative position angles exceeding $| 90^\circ |$  is
large compared to the sample studied by \citet{vanBuren95}. 
Because our sample was selected on the basis of IR
morphology rather than  on the basis of stars known to have
high proper motion \citep[e.g.,][]{vanBuren95, Peri15}, it
is not surprising that our sample displays a broader
distribution of relative position angles. The broad
distribution of relative position angles may indicate that
factors other than a star's peculiar motion could produce
relative velocities between a star and the surrounding
interstellar material that generate  a bowshock nebula. 
Section~\ref{env.sec} above discusses physical scenarios
that could explain these  objects.

\section{Conclusions}

We have compiled the largest collection to date of candidate
stellar bowshock nebulae selected on the basis of mid-IR
morphology at 24 or 22 $\mu$m from \sst\ and $WISE$ sky
surveys of the Galactic Plane.  The vast majority ---
\numnewhere\ of the \numtot\ objects --- are identified here
for the first time.  A minority of objects ($\sim$19\%)
appear at 8 $\mu$m or shorter wavelengths as well,
indicating a population of very small hot dust  grains
within the swept-up nebula, or possibly a PAH contribution. 
While the infrared images alone do not demonstrate that the
tabulated objects are necessarily shocks, the preponderance
of ancillary evidence, including proper motions and spectral
classifications  of the host stars, strongly suggests that
this is the best interpretation for the majority of
objects.  On further investigation, some of the objects may turn
out to be other phenomena, such as asymmetric dust shells
around evolved stars. 

The distribution of these objects on the sky is tightly
confined to the Galactic mid-Plane, consistent with their
production by massive stars which have strong stellar winds.
Proper motions of the central stars, where known, indicate a
clear excess of objects having velocity vectors  aligned
with the symmetry axis of the infrared nebula.  This is
consistent with their production by a population of massive
runaway stars.   While the majority of the candidate 
bowshock nebulae and their central stars lie in relatively
isolated environments, a substantial subset ($\sim$20\%)
either face giant \hii\ regions or face bright-rimmed
clouds.  These may, respectively, constitute two classes of
``in-situ'' bowshocks where either an external flow 
overruns the star or ``photoevaporative flow'' bowshocks
where a stellar wind interacts with material evaporating
from a nearby molecular cloud skin. The correlation of
bowshock orientations on angular scales of
$\lesssim$10\arcmin\ provides evidence that local
environmental phenomena, such as outflows from star forming
regions, produce the relative motions between a nearby
massive star and the ISM  that generate the bowshock
nebulae. 

Table~\ref{bigtable.tab} and the Appendix of \numtot\
stellar bowshock candidates may be used to further
investigate the  origins of runaway stars, phenomena
associated with large-scale outflows of gas from star
forming regions, and the physics of massive star winds.  
Additional scrutiny of existing  infrared surveys could be
expected to find a few additional  bowshock candidates,
especially at higher Galactic latitudes not covered by our
search.  However, we expect the yield to be small as the
areal density of bowshocks within the \sst\ Plane surveys
drops rapidly with latitude.  In a forthcoming series of
papers  we will present an analysis of the characteristics
of this comprehensive bowshocks sample, including their 
spectral types, binarity, distances, spectral energy distributions, and local
environments.

\acknowledgements Our team is grateful to support from the
National Science Foundation through grant AST-1412845,
AST-1411851, and REU grant AST-1063146, as well as NASA
through grant NNX14AR35A. We thank our anonymous reviewer
for comments that helped clarify aspects of this manuscript.

\clearpage

\begin{deluxetable}{ cccccccccccccc}
\tablecaption{Central Stars of Bowshock Nebula Candidates \label{bigtable.tab}}
\tablecolumns{14}
\floattable
\tablewidth{10.0in}
\tabletypesize{\scriptsize} 
\rotate
\tablehead{ \colhead{Num} & \colhead{Name } & \colhead{R.A. }     & \colhead{Decl. }   & \colhead{Ref.} & \colhead{Alias} & \colhead{8.0?} & \colhead{Flag} & \colhead{R$_0$ }   & \colhead{P.A. } & \colhead{H}      & \colhead{I2/W2}   & \colhead{A$_K$}  & \colhead{Env.} \\
            \colhead{}    & \colhead{}      & \colhead{(2000.0) } & \colhead{(2000.0)} & \colhead{}     & \colhead{}      & \colhead{}     & \colhead{}     & \colhead{(arcsec)} & \colhead{(deg)} & \colhead{(mag)} & \colhead{(mag)} & \colhead{(mag)} & \colhead{} 
}
\startdata
1 & G000.1169$-$00.5703 & 17:48:07.70 & $-$29:07:55.5 & T &  & Y & C & 26.4 & 27 & 8.97 & 8.72 & 0.22 & I \\
2 & G000.3100$-$01.0495 & 17:50:27.59 & $-$29:12:46.8 & T &  & N & C & 6.8 & 145 & 10.57 & 10.31 & 0.23 & I \\
3 & G001.0563$-$00.1499 & 17:48:41.78 & $-$28:06:37.8 & T &  & Y &   & 6.2 & 200 & 11.14 & 9.8 & 1.22 & H \\
4 & G001.2588$-$00.0780 & 17:48:53.40 & $-$27:53:59.7 & T &  & N & C & 10.2 & 215 & 12.28 & 11.22 & 0.96 & FH \\
5 & G003.5118$-$00.0470 & 17:53:56.10 & $-$25:56:50.3 & T &  & N &   & 8.6 & 135 & 10.69 & 9.69 & 0.91 & I \\
6 & G003.7391+00.1425 & 17:53:43.26 & $-$25:39:19.2 & T &  & N &  & 3.9 & 20 & 11.48 & 9.41 & 1.89 & I \\
7 & G003.8417$-$01.0440 & 17:58:30.64 & $-$26:09:49.1 & T &  & Y &  & 12.6& 25 & 10.22 & 9.87 & 0.31 & FB \\
8 & G004.3087+00.2222 & 17:54:41.61 & $-$25:07:25.6 & T &  & N &  & 10.1 & 260 & 13.94 & 12.1 & 1.68 & I \\
9 & G004.7315$-$00.3875 & 17:57:57.62 & $-$25:03:53.6 & T &  & N & C & 10.1 & 125 & 13.3 & 11.77 & 1.39 & I \\
10& G004.8449$-$00.9309 & 18:00:17.51 & $-$25:14:14.7 & T & HD314937 & N &  & 3.1 & 140 & 9.36 & 9.31 & 0.03 & I \\
11 & G005.5941+00.7335 & 17:55:36.25 & $-$23:45:21.8 & T &  & N &   & 7.6 & 350 & 9.78 & 9.22 & 0.50 & I \\
12 & G005.6985$-$00.6350 & 18:01:06.15 & $-$24:21:34.4 & T &  & N &   & 20.8 & 180 & 11.56 & 9.11 & 2.24 & I \\
13 & G006.2812+23.5877 & 16:37:09.54  & $-$10:34:01.5  & V88\tablenotemark{a} & Zeta Oph & - &  & 29 & 30 & 2.59 & -99 & -99 & I \\
14 & G006.2977$-$00.2012 & 18:00:40.59 & $-$23:36:50.7 & T &  & N & C & 24 & 210 & 13.07 & 12.16 & 0.82 & H \\
15 & G006.3600$-$00.1846 & 18:00:44.92 & $-$23:33:06.3 & T &  & N &  & 5.5 & 110 & 9.32 & 9.11 & 0.18 & I \\
16 & G006.8933+00.0743 & 18:00:55.27 & $-$22:57:36.9 & T &  & N & C & 29 & 65 & 8.73 & 8.36 & 0.33 & I \\
17 & G007.5265$-$00.2652 & 18:03:33.49 & $-$22:34:38.3 & T &  & N &  & 4.1 & 350 & 9.06 & 8.11 & 0.86 & I \\
18 & G008.3690+00.0239 & 18:04:15.53 & $-$21:42:05.3 & T &  & N & C & 27 & 50 & 9.07 & 8.3 & 0.69 & I \\
19 & G009.0177+00.1410 & 18:05:11.21 & $-$21:04:43.3 & T &  & N &  & 9.1 & 290 & 8.85 & 8.41 & 0.39 & I \\
20 & G009.6852$-$00.2025 & 18:07:51.86 & $-$20:39:48.8 & T &  & N &  & 8.3 & 335 & 10.92 & 9.69 & 1.12 & I 
\enddata
\tablenotetext{a}{\citep{vanBuren88}}
\tablecomments{The first 20 bowshock nebulae candidates. 
The complete list of \numtot\ candidates is available as a machine readable table 
in the electronic version of the Journal. (1) Numeration of this work, 2(Generic name in Galactic coordinates,
(3) Right ascension, (4) Declination, (5) First reported discovery reference; T denotes This Work, (6) Alternate name for star, (7)
Flag indicating whether source is detected at 8 $\mu$m, 
(8) C denotes that there are multiple candidates for the central star, (9) Angular distance between the
star and nebular apex in arcsec, (10) Position angle of nebula in degrees, (11) $H$-band magnitude, (12) IRAC 4.5 $\mu$m or WISE 4.6 $\mu$m 
magnitude, (13) Estimated extinction, (14) Code indicating the local environment, as described in
Section~2.2}
\end{deluxetable}

\clearpage

\clearpage

\clearpage

\clearpage

\clearpage

\clearpage

\clearpage

\clearpage

\clearpage

\clearpage

\clearpage

\clearpage

\appendix  \section{Atlas of Images} This electronic-only
appendix contains an atlas of three-color images for each of
the \numtot\ bowshock nebula candidates.  For the majority
of the candidates that have \sst\ data, the  color scheme is
24/8.0/4.5 $\mu$m in red/green/blue.    For objects outside
the coverage of the \sst\ surveys, we use the $WISE$
22/12/4.6 bands in red/green/blue.  In a few cases where the
\sst\ 24 $\mu$m data are not available, we substitute the
$WISE$ 22 $\mu$m data.  Images are  2\arcmin\ square unless
a larger field of view is required to show the entire
nebulae or pertinent surrounding features.  White arrows on some images
depict the direction and magnitude of central star's proper motion, where known.

\end{document}